\documentclass[twoside]{LCWS11}
\usepackage[latin1]{inputenc}
\usepackage[dvips]{graphicx, epsfig, color}
\usepackage{wrapfig, rotating}
\usepackage{amssymb, amsmath, array, slashed}
\usepackage{cite}
\begin{document}
\title{Testing Little Higgs Mechanism at Future Colliders}
\author{Shigeki Matsumoto
\vspace{.3cm}\\
IPMU, TODIAS, University of Tokyo \\
Kashiwa-no-ha 277-8583, Kashiwa, Japan}

\maketitle

\begin{abstract}
In the little higgs scenario, several coupling constants are related to each other to guarantee the stability of the higgs boson mass at one-loop level. This relation is called the little higgs mechanism. We discuss how accurately the relation can be tested at future $e^+e^-$ colliders, with especially focusing on the top sector of the scenario using a method of effective lagrangian. In order to test the mechanism, it is important to measure the Yukawa coupling of the top partner. Higgs associated and threshold productions of the top partner are found to be the best processes for this purpose.
\end{abstract}

%%%%%%%%%%%%%%%%%%%%%%%%
%%%%% Introduction %%%%%
%%%%%%%%%%%%%%%%%%%%%%%%
\section{Introduction}

The standard model (SM) is well known to have the little hierarchy problem~\cite{little_hierarchy}, which is essentially from quadratically divergent corrections to the higgs mass term. The little higgs scenario~\cite{LH} has been proposed to solve the problem, where the higgs boson is regarded as a pseudo Nambu-Goldstone boson associated with the breaking of a global symmetry at the ${\cal O}(10)$ TeV scale. Explicit breaking terms of the symmetry are specially arranged to cancel the corrections at one-loop level. The mechanism of this cancellation is called the little higgs mechanism, which is commonly equipped in all models of the scenario.

The little higgs mechanism predicts the existence of new particles called little higgs partners at the ${\cal O}(1)$ TeV scale. The mechanism also predicts some relations between coupling constants of SM interactions and those of the new particles. Among the partners, the top partner is the most important one, because it is responsible for the cancelation of the largest quadratically divergent correction. Though the top partner has a color charge and could be produced at the LHC, its discovery does not mean the confirmation of the little higgs scenario. This is because new particles which are similar to the top partner are also predicted in various new physics models. In order to test the little higgs scenario, we have to verify the relation between interactions predicted by the little higgs mechanism.

This verification requires us to measure the Yukawa coupling of the top partner. Future linear colliders such as the international linear collider (ILC)~\cite{ILC} and the compact linear collider (CLIC)~\cite{CLIC} give a good opportunity for coupling measurements. Following four processes are considered in this report; higgs associated productions, $e^+e^- \to T \bar{T} h$, $t \bar{T} h + T \bar{t} h$, and threshold productions, $e^+e^- \to T \bar{T}$, $T\bar{t} + t\bar{T}$, where $t$, $T$, and $h$ are top quark, top partner, and higgs boson, respectively. We found that the coupling can be measured precisely using the associate production, $e^+e^- \to T \bar{T} h$, when the center of mass energy is large enough. The threshold production, $e^+e^- \to T \bar{T}$, also allows us to measure it with the same precision. Interestingly, with smaller center of mass energy, it is even possible to measure the coupling using the threshold production, $e^+e^- \to \bar{t} T + T \bar{t}$.

In the following, after introducing the effective lagrangian to describe the top sector, higgs associated and threshold productions of the top partner are discussed with focusing on how these cross sections are sensitive to the Yukawa coupling of the top partner. Since this report is based on Ref.~\cite{Harigaya:2011yg}, please see the reference for more details.

%%%%%%%%%%%%%%%%%%
%%%%% Set up %%%%%
%%%%%%%%%%%%%%%%%%
\section{Top sector of the little higgs scenario}

\begin{figure}[t]
\centerline{\includegraphics[width=0.4\columnwidth]{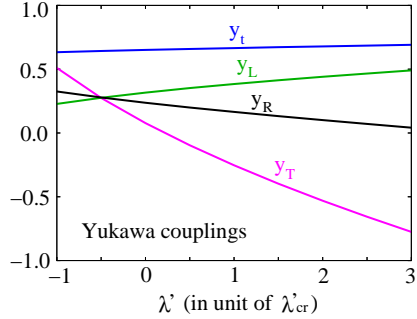}}
\caption{\small Yukawa couplings, $y_t$, $y_T$, $y_L$, $y_R$, as a function of $\lambda'$ (in unit of $\lambda'_{cr}$).}
\label{fig: yukawas}
\end{figure}

Vector-like quark called the top partner is necessarily introduced in the little higgs scenario, and its interactions with higgs and top quark are described by the effective lagrangian,
\begin{eqnarray}
{\cal L}_{\rm eff}
&=&
- m_U \bar{U}_L U_R
- y_3 \bar{Q}_{3L} H^c u_{3R}
- y_U \bar{Q}_{3L} H^c U_R
\nonumber\\
&&
- (\lambda /m_U) \bar{U}_L u_{3R} |H|^2
- (\lambda'/m_U) \bar{U}_L U_R |H|^2 + h.c.,
\label{eq: effective lagrangian}
\end{eqnarray}
where $Q_{3L}=(u_{3L},b_{3L})^T$ and $u_{3R}$ are third generation left- and right-handed quarks of the SM, while $U_L$ and $U_R$ are left- and right-handed top partners whose quantum numbers are the same as $u_{3R}$. Higgs boson is denoted by $H^c$ with '$c$' meaning charge conjugation. Since quadratically divergent corrections to the higgs mass term should be cancelled at 1-loop level, the following relation between the coupling constants, $y_3$, $y_U$, and $\lambda^\prime$, is required, $-2\lambda' = y_3^2 + y_U^2$, which is nothing but the little higgs mechanism at the top sector.

Once the electroweak symmetry is broken, third generation quarks are mixed with top partners. In following discussions, we use the notations; $m_t$ ($m_T$) is the mass of the top quark $t$ (top partner $T$) and $\sin \theta_L$ ($\sin \theta_R$) is the mixing angle between their left-handed (right-handed) components. Though the model parameters, $m_U$, $y_3$, $y_U$, $\lambda$, $\lambda^\prime$, are originally used in the effective lagrangian, we use following five parameters, $m_t$, $m_T$, $\sin\theta_L$, $\lambda$, $\lambda^\prime$, as fundamental ones. Other parameters, $m_U$, $y_3$, $y_U$, $\tan\theta_R$, are therefore given as functions of the fundamental parameters. Yukawa interactions of $t$ and $T$ are given by
\begin{eqnarray}
{\cal L}_{\rm yukawa}
=
- y_t \bar{t} t h
- y_T \bar{T} T h
- (\bar{T} [y_L P_L + y_R P_R] th + h.c.),
\label{eq: interactions}
\end{eqnarray}
where $h$ is the higgs field. There are also gauge interactions of $t$ and $T$ (see Ref.\cite{Harigaya:2011yg} for their concrete expressions). Yukawa couplings are approximately given by $y_t \simeq m_t/v$, $y_T \simeq m_T \sin^2 \theta_L/v + \lambda' v \cos \theta_L /m_T$, $y_L \simeq m_T \sin \theta_L/v$, and $y_R \simeq m_t \sin \theta_L/v + \lambda v/m_T$.

The little higgs mechanism is described by the relation between $y_3$, $y_U$, and $\lambda^\prime$. First two parameters $y_3$ and $y_U$ are almost determined by $m_t$, $m_T$, and $\sin \theta_L$, which can be measured precisely when $T$ is discovered. On the other hand, we have to measure $y_T$ to determine the last parameter $\lambda^\prime$. We therefore fix these parameters using the representative point,
\begin{eqnarray}
m_T = 400 {\rm GeV},
\qquad
\sin \theta_L = 0.2,
\qquad
{\rm Br}(T \rightarrow t h)/{\rm Br}(T \rightarrow b W) = 0.98,
\label{eq: representative point}
\end{eqnarray}
where the ratio of branching fractions corresponds to the one obtained by assuming $\lambda = 0$ with keeping the relation of the little higgs mechanism. Higgs mass is fixed to be $m_h = 120$ GeV. The point satisfies all phenomenological constraints and is also attractive from the viewpoint of naturalness on the little hierarchy problem. Since $m_t$ has already been measured precisely, there are four free parameters in the effective lagrangian. It is therefore possible to test the little higgs mechanism by measuring one more observable, which should be sensitive to $\lambda'$. As can be seen in Fig.~\ref{fig: yukawas}, in which Yukawa couplings are shown as functions of $\lambda'$ in unit of $\lambda^\prime_{cr} \equiv -(y_3^2 + y_U^2)/2$ with other model parameters being fixed to be those satisfying the conditions of the representative point, $y_T$ is the most sensitive against the change of $\lambda'$ as expected. We thus focus on physical quantities involving this Yukawa coupling.

%%%%%%%%%%%%%%%%%%%%%%%%%%%%%%%%%
%%%%% Associate productions %%%%%
%%%%%%%%%%%%%%%%%%%%%%%%%%%%%%%%%
\section{Associate Productions}
\label{sec: associate}

\begin{figure}[t]
\centerline{
\includegraphics[width=0.45\columnwidth]{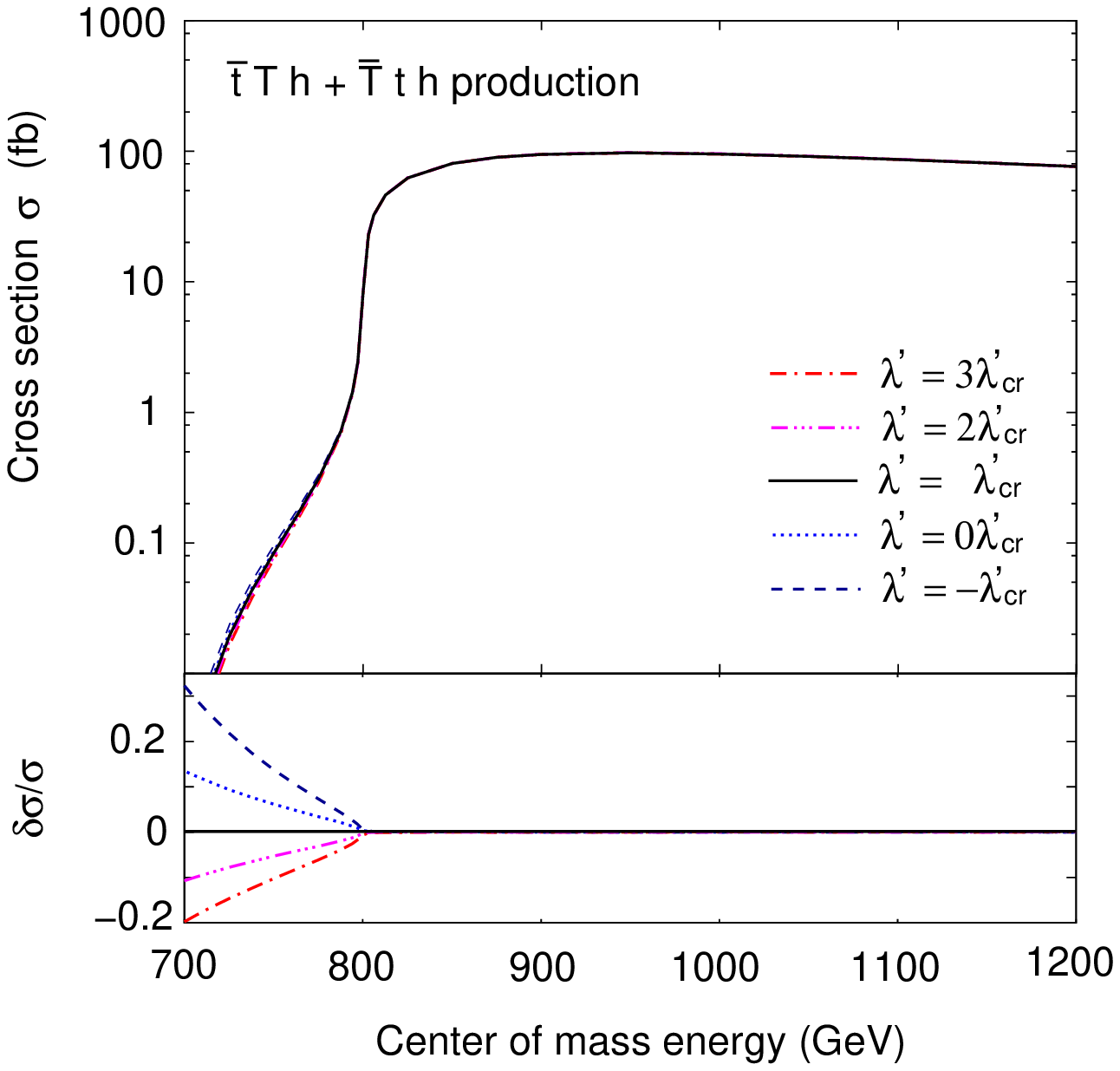}
~
\includegraphics[width=0.45\columnwidth]{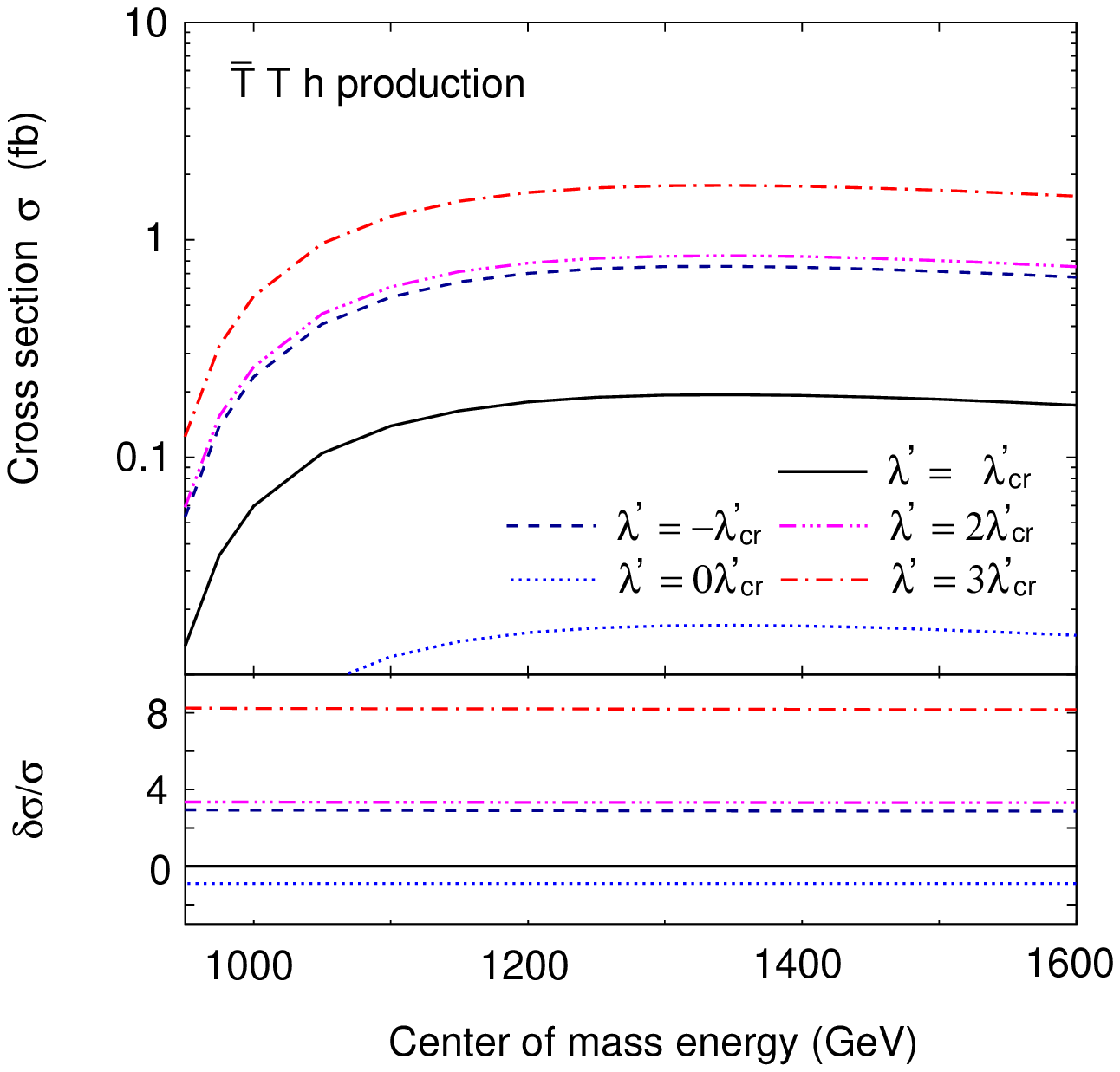}}
\caption{\small Cross sections for higgs associated productions $e^+ e^- \to t \bar{T} h + T \bar{t} h$ and $e^+ e^- \to T \bar{T} h$.}
\label{fig: associate productions}
\end{figure}

Top partner productions associated with a higgs boson enable us to determine $y_T$ by measuring their cross sections with an appropriate center of mass energy. Two higgs associated processes exist. One is $e^+ e^- \to t \bar{T} h + T \bar{t} h$, and another is $e^+ e^- \to T \bar{T} h$. The sum of the cross sections, $\sigma(e^+ e^- \to t \bar{T} h) + \sigma(e^+ e^- \to T \bar{t} h)$, are shown in Fig.\ref{fig: associate productions} (upper part of the left panel) as a function of center of mass energy with several choices of $\lambda^\prime$. Other model parameters are fixed according to the conditions in eq.(\ref{eq: representative point}). In order to see the sensitivity of the cross section against the change of $\lambda^\prime$, we also plot the deviation of the cross section from the one predicted by the little higgs mechanism, $\delta \sigma/\sigma \equiv [\sigma(\lambda^\prime) - \sigma(\lambda^\prime_{cr})] / \sigma(\lambda^\prime_{cr})$ (lower part of the left panel). The deviation becomes almost zero when the center of mass energy exceeds 800 GeV, because the production of two on-shell top partners dominates the associate production, whose production cross section and branching ratio of $T \to th$ are independent of $\lambda^\prime$. When the center of mass energy is below 800 GeV, the cross section is as small as 0.1 fb and the determination of $\lambda^\prime$ seems to be difficult in this associate production.

On the other hand, another associate production, $e^+e^-\rightarrow T\bar{T}h$, is interesting, whose cross section and its deviation from the little higgs prediction are shown in Fig.\ref{fig: associate productions} (right panel). The cross section depends strongly on the value of $\lambda^\prime$, which enable us to explore the little higgs mechanism accurately using this association process, though the production cross section itself is not so large and the center of mass energy must be large enough.

%%%%%%%%%%%%%%%%%%%%%%%%%%%%%%%%%
%%%%% Threshold productions %%%%%
%%%%%%%%%%%%%%%%%%%%%%%%%%%%%%%%%
\section{Threshold productions}
\label{sec: threshold}

\begin{figure}[t]
\centerline{
\includegraphics[width=0.45\columnwidth]{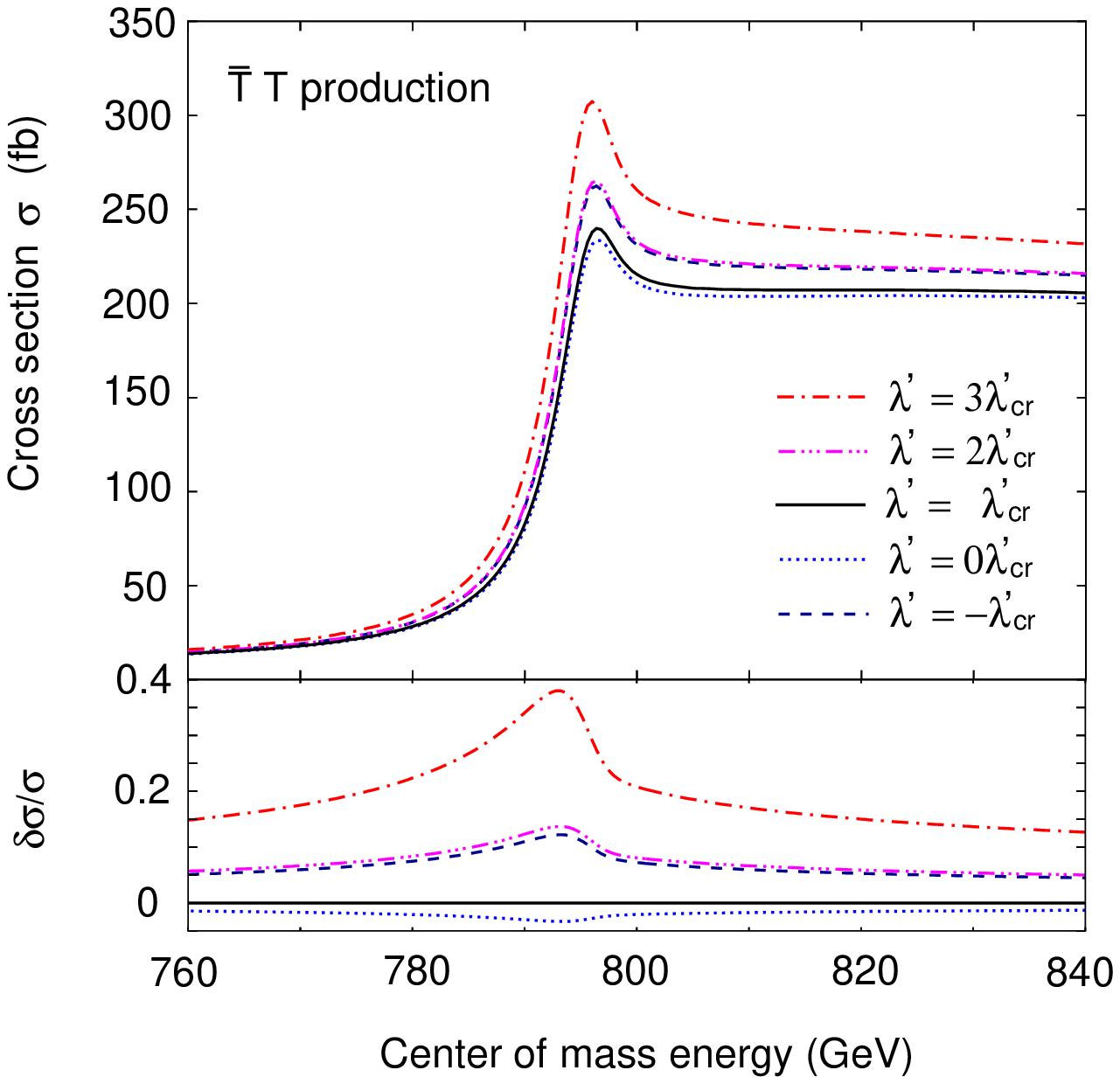}
~
\includegraphics[width=0.45\columnwidth]{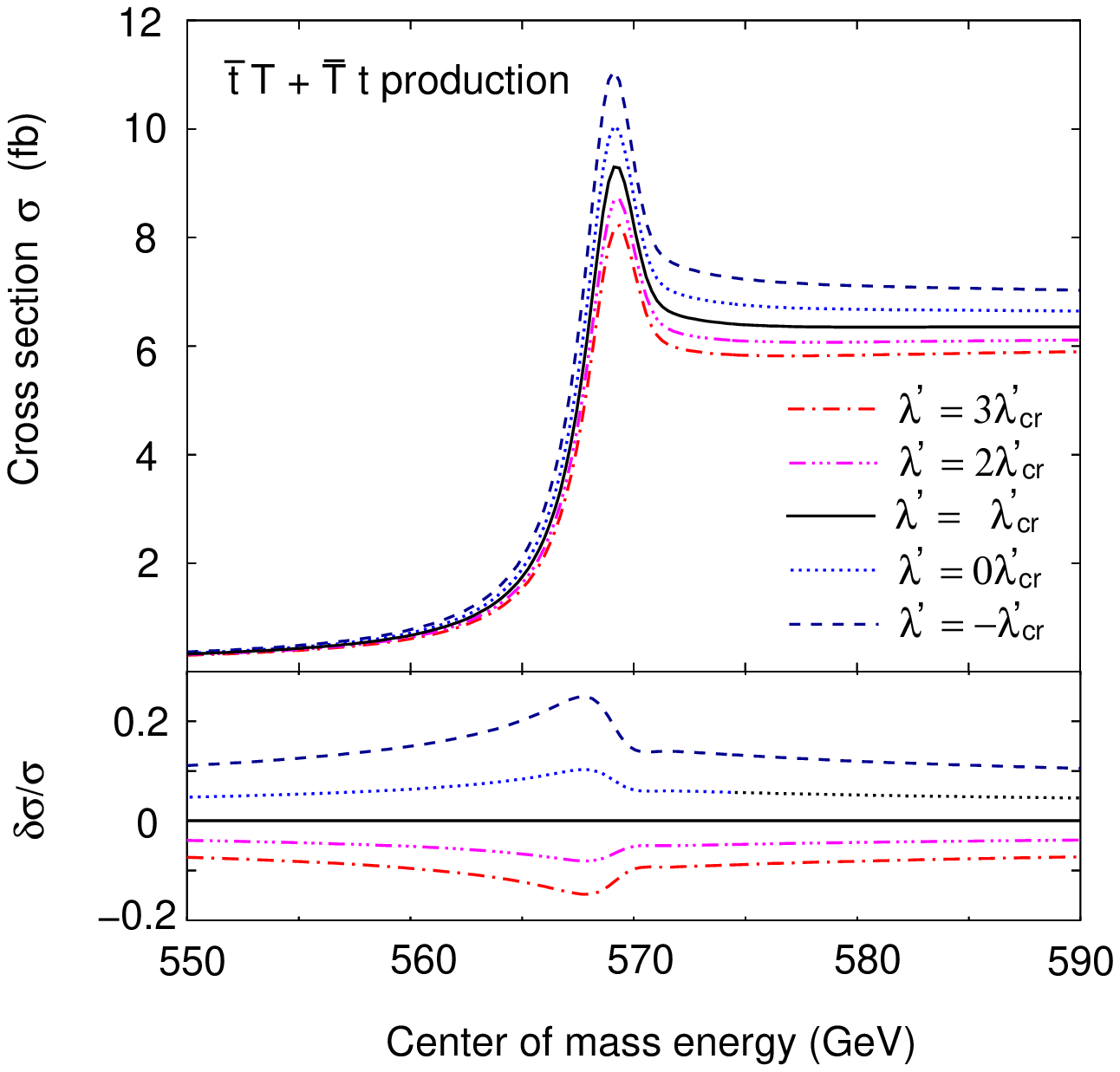}}
\caption{\small Cross sections for threshold productions $e^+ e^- \to T \bar{T} $ and $e^+ e^- \to t \bar{T} + T \bar{t}$.}
\label{fig: threshold productions}
\end{figure}

We next consider top partner productions, $e^+ e^- \to T \bar{T}$ and $ e^+ e^- \to t \bar{T} + T \bar{t}$, at their threshold energies. Since the cross sections of these processes are significantly affected by the exchange of virtual higgs bosons due to the threshold singularity~\cite{Strassler:1990nw}, the Yukawa coupling $y_T$ is expected to be determined precisely. The production cross section of the top partner pair at threshold energy is shown in the left panel of Fig.~\ref{fig: threshold productions} with several choices of $\lambda^\prime$. The deviation of the cross section from the prediction of the little higgs mechanism is also shown in this figure as in the case of higgs associated productions. It can be seen that the deviation becomes maximum at the peak of the cross section, which corresponds to the first bound state composed of top partners. Since the cross section is huge at the peak, which is about 250 fb$^{-1}$, the Yukawa coupling $y_T$ is expected to be determined precisely. It should be noticed that the cross section does not depend on the sign of $y_T$, so that there is two-fold ambiguity in the determination of $y_T$, as can be seen in the figure. This ambiguity is resolved by investigating other production channels such as higgs associated productions.

Production cross section of a top quark and a top partner at threshold energy as well as its deviation from the little higgs prediction are plotted in the right panel of Fig.~\ref{fig: threshold productions}. The deviation is again maximized at the peak corresponding to the first bound state. Though the cross section is lower than that of the top partner pair production, it is still possible to determine $\lambda^\prime$ precisely if the integrated luminosity is large enough. One advantage is that the determination of $\lambda^\prime$ is possible even if the center of mass energy is not so large.

%%%%%%%%%%%%%%%%%%%%%%
%%%%% Capability %%%%%
%%%%%%%%%%%%%%%%%%%%%%
\section{Testing the little higgs mechanism}
\label{sec: detectability}

\begin{figure}[t]
\centerline{\includegraphics[width=0.4\columnwidth]{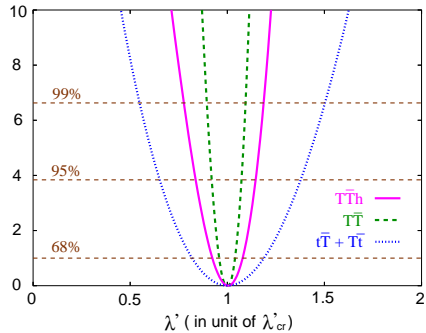}}
\caption{\small The $\chi^2$ distributions for $e^+ e^- \to T \bar{T} h$, $T \bar{T}$, and $t \bar{T} + T \bar{t}$ processes.}
\label{fig: chi2}
\end{figure}

We now consider the capability of future linear colliders to test the little higgs mechanism. Testing the mechanism is equivalent to the determination of $\lambda^\prime$, so that we discuss how accurately $\lambda^\prime$ can be determined at collider experiments using the processes discussed in previous sections. We focus on the processes $e^+ e^- \to T \bar{T} h$, $e^+ e^- \to T \bar{T}$, and $e^+ e^- \to t \bar{T} + T \bar{t}$, because these processes are very sensitive to $\lambda^\prime$. The center of mass energy are set to be 1000 GeV, 794 GeV and 568.4 GeV for $T \bar{T} h$, $T \bar{T}$ and $t \bar{T} + T \bar{t}$ productions, respectively, in order to maximize the capability. On the other hand, the integrated luminosity is fixed to be ${\cal L}_{\rm eff} = 500$ fb$^{-1}$ in each process, where ${\cal L}_{\rm eff}$ is the effective luminosity defined by ${\cal L}_{\rm eff} = {\cal E} \times {\cal L}$ with ${\cal E}$ and ${\cal L}$ being the efficiency factor and the original integrated luminosity. The factor depends on the acceptance of collider detectors and kinematical cuts used to reduce backgrounds from SM processes, which is determined precisely when experiments start. Main background against the signal is from the SM process of top quark production. Since the cross section of the SM process is not too large compared to those of signal processes, background reductions are expected to be performed efficiently by imposing appropriate kinematical cuts. The efficiency factor from background reductions can be estimated using Monte-Carlo simulations. In our calculation, however, we take the efficiency factor so that the effective luminosity becomes 500 fb$^{-1}$ with simply assuming efficient background reductions. We remain the detailed calculation of the factor as a future problem.

With the use of the effective luminosity ${\cal L}_{\rm eff}$, the $\chi^2$ function, which quantifies how accurately the determination of $\lambda^\prime$ can be performed at collider experiments, is defined by
\begin{eqnarray}
\chi^2 (\lambda^\prime)
\equiv
\frac{\left[N(\lambda') - N(\lambda'_{cr})\right]^2}{N(\lambda'_{cr})}
=
{\cal L}_{\rm eff}
\times
\frac{\left[\sigma(\lambda') - \sigma(\lambda'_{cr})\right]^2}
{\sigma(\lambda'_{cr})},
\end{eqnarray}
where $N(\lambda^\prime) = {\cal L}_{\rm eff} \sigma(\lambda^\prime)$ is the number of the signal event with fixed $\lambda^\prime$, which will be obtained at collider experiments after imposing kinematical cuts and considering detector acceptances. The resultant $\chi^2$ distributions for the higgs associated production, $e^+e^- \to T \bar{T} h$, and threshold productions, $e^+e^- \to T \bar{T}$ and $e^*e^- \to t \bar{T} + T \bar{t}$, are shown in Fig.~\ref{fig: chi2}, where the center of mass energy is fixed to be 1000 GeV, 794 GeV and 568.4 GeV, respectively. It can be seen that the coupling constant $\lambda^\prime$ can be measured with 8\% accuracy using the higgs associated production. If the center of mass energy is possible to be increased to 1350 GeV, the coupling $\lambda^\prime$ can be measured with 4\% accuracy using the same process.

On the other hand, it can be seen that the coupling constant $\lambda^\prime$ can be determined with 4\% accuracy using the threshold production of top partner pair production. In this case, the center of mass energy required for the determination is smaller than that of the higgs associated production. Furthermore, using the threshold production of a top quark and a top partner, the coupling constant $\lambda^\prime$ can be measured with 20\% accuracy even if the center of mass energy is smaller than other processes , which is around 500 GeV.

%%%%%%%%%%%%%%%%%%%
%%%%% Summary %%%%%
%%%%%%%%%%%%%%%%%%%
\section{Summary}
\label{sec: summary}

We have studied the capability of future collider experiments to test the little higgs mechanism. The mechanism predicts a certain relation between coupling constants in the top sector of the little higgs scenario. The test of the mechanism is essentially equivalent to the measurement of the Yukawa coupling of the top partner, namely the coupling constant $\lambda^\prime$. It is therefore very important to investigate how accurately $\lambda^\prime$ can be measured at the experiments. With the use of an appropriate representative point, we found that the coupling constant can be measured with a few percent accuracy using the higgs associated production, $e^+e^- \to T \bar{T} h$, and the threshold production, $e^+e^- \to T \bar{T}$, when the center of mass energy is ${\cal O}(1)$ TeV. On the other hand, using the threshold production, $T\bar{t} + t\bar{T}$, the measurement of $\lambda^\prime$ is still possible with a few ten percent accuracy even if the center of mass energy is around 500 GeV. When the top partner is discovered at the LHC experiment, the processes emphasized in this report will be important to confirm whether the top partner is really the one predicted by the little higgs scenario or not.

\section{Acknowledgments}

This work is supported by Grant-in-Aid for Scientific research from the Ministry of Education, Science, Sports, and Culture (MEXT), Japan, (Nos.\ 22244021 \& 23740169) and World Premier International Research Center Initiative (WPI Initiative), MEXT, Japan.

\begin{footnotesize}

\end{footnotesize}


\begin{thebibliography}{99}

\bibitem{little_hierarchy}
For recent analysis, see
%\bibitem{Barbieri:2004qk}
R.~Barbieri, A.~Pomarol, R.~Rattazzi, A.~Strumia,
%``Electroweak symmetry breaking after LEP-1 and LEP-2,''
Nucl.\ Phys.\  {\bf B703}, 127-146 (2004);
%[hep-ph/0405040]
%\bibitem{Han:2004az}
Z.~Han, W.~Skiba,
%``Effective theory analysis of precision electroweak data,''
Phys.\ Rev.\  {\bf D71}, 075009 (2005).
%[hep-ph/0412166]

\bibitem{LH}
%\bibitem{ArkaniHamed:2001nc}
N.~Arkani-Hamed, A.~G.~Cohen, H.~Georgi,
%``Electroweak symmetry breaking from dimensional deconstruction,''
Phys.\ Lett.\  {\bf B513}, 232-240 (2001);
%[hep-ph/0105239]
%\bibitem{Perelstein:2005ka}
and, for a review, see
M.~Perelstein,
%``Little Higgs models and their phenomenology,''
Prog.\ Part.\ Nucl.\ Phys.\  {\bf 58}, 247 (2007).
%[arXiv:hep-ph/0512128].

\bibitem{ILC}
\verb$http://www.linearcollider.org/$.

\bibitem{CLIC}
\verb$http://clic-study.web.cern.ch/clic-study/$.

\bibitem{Harigaya:2011yg}
K.~Harigaya, S.~Matsumoto, M.~M.~Nojiri and K.~Tobioka,
%``Testing Little Higgs Mechanism at Future Colliders,''
JHEP {\bf 1201}, 135 (2012).
%[arXiv:1109.4847 [hep-ph]].

\bibitem{Strassler:1990nw}
M.~J.~Strassler, M.~E.~Peskin,
%``The Heavy top quark threshold: QCD and the Higgs,''
Phys.\ Rev.\  {\bf D43}, 1500-1514 (1991).

\end{thebibliography}
\end{document}